\documentclass[fleqn,usenatbib]{mnras}

\usepackage{newtxtext,newtxmath}

\usepackage[T1]{fontenc}
\usepackage{ae,aecompl}

\usepackage{graphicx}	
\usepackage{amsmath}	
\usepackage{bm}
\usepackage{xspace}

\usepackage{todonotes}
\newcommand{\bhac}{\texttt{BHAC}~}

\usepackage[normalem]{ulem}

\title[Magnetic field structure in Sgr A*]{Magnetic field structure in
the vicinity of a super-massive black hole in low luminosity galaxies:
the case of Sgr A*}

\author[Nathanail, Dhang \& Fromm]{Antonios Nathanail$^{1,5}$ \thanks{E-mail:
    nathanail@itp.uni-frankfurt.de},  Prasun
Dhang$^{2,3}$ and Christian M. Fromm$^{4,5,6}$ \\
$^{1}$ Department of Physics, National and Kapodistrian University of
   Athens, Panepistimiopolis, GR 15783 Zografos, Greece   \\
$^{2}$Institute for Advanced Study, Tsinghua University, Beijing-100084, China\\
$^{3}$Department of Astronomy, Tsinghua University, Beijing-100084, China   \\
$^{4}$ Institut f\"ur Theoretische Physik und Astrophysik,
Universit\"at W\"urzburg, Emil-Fischer-Strasse 31, 97074 W\"urzburg,
Germany \\
$^{5}$Institut f\"ur Theoretische Physik, Goethe Universit\"at
Frankfurt, Max-von-Laue-Str.1, 60438 Frankfurt am Main, Germany \\
$^{6}$Max-Planck-Institut f\"ur Radioastronomie, Auf dem H\"ugel 69,
D-53121 Bonn, Germany   }

\usepackage{hyperref}
\hypersetup{
  linkcolor=cyan,         
  citecolor=blue,      
  filecolor=magenta,      
  urlcolor=magenta            
}

\begin{document} 
\label{firstpage}

\maketitle

\begin{abstract}

	Observations of $\rm SgrA^*$ have provided a lot of insight  on
	low-luminosity accretion, with a handful of bright flares
	accompanied with orbital motion close to the horizon. It has been
	proposed that gas supply comes from stellar winds in the
	neighborhood of the supermassive black hole. We here argue that
	the flow at the vicinity of the black hole has a  low
	magnetization and  a structure of alternating polarity totally
	dictated by the well studied and long-ago proposed MRI turbulent
	process. This can be the case, provided that in larger distances
	from the black hole magnetic diffusivity is dominant and thus the
	magnetic field will never reach equipartition values.  For  $\rm
	SgrA^*$, we show the immediate consequences of this specific
	magnetic field geometry, which are: (i) an intermittent flow that
	passes from quiescent states to flaring activity, (ii) no
	quasi-steady-state jet, (iii) no possibility of a magnetically
	arrested configuration. Moreover a further distinctive feature of
	this geometry is the intense magnetic reconnection events,
	occurring as layers of opposite magnetic polarity are accreted,
	in the vicinity of the black hole. Finally, we argue that the absence 
	of a jet structure in such case will be a smoking gun in 43 \& 86 GHz
	observations.

\end{abstract}

\begin{keywords}
{Black hole physics, Sgr A*, Magnetohydrodynamics}
\end{keywords}



\section{Introduction}
\label{sec:intro}

The supermassive black hole (BH) in the Galactic center, Sgr A*, is the
closest of its kind and is the subject of several campaigns of
multi-wavelength observations, since it is an excellent probe for
accretion physics in the extreme gravity regime. Over the years
unparalleled insight on the accretion flow in the immediate environment
of the central BH has been obtained
\citep{Falcke1998,Doeleman2008}.
Sgr A* is one of the
primary sources for the Event Horizon Telescope (EHT) and GRAVITY
collaborations yielding extraordinary results the last years and more
expected to come \citep{Johnson2015,Abuter2018b}. 

Multi-wavelength observations have revealed a low accretion rate with an
extremely low bolometric luminosity $L_{\rm bol}\sim 10^{-9}L_{\rm Edd}$,
where $L_{\rm Edd}$ is the Eddington Luminosity \citep{Baganoff2003,
Bower2019}, which places the accretion flow of $\rm SgrA^*$ in the regime
of a radiatively inefficient accretion flow \citep{Yuan2014}.

Observations have hinted that around 80 massive stars reside in a
rotating disk inside the parsec neighborhood of the Galactic
center\citep{Martins2007}.
It has been suggested that 30 of
these massive stars, identified as Wolf-Rayet (WR) magnetized stars could
source the accretion flow around the Galactic center
\citep{Quataert2004,Loeb2004,Shcherbakov2009, Ressler2020b}.
This scenario has been studied in  a multi-scale simulation
\citep{Ressler2020} and suggested that the accretion flow around Sgr A*
is in a magnetically arrested (MAD) state, moreover, its dynamics
differ significantly from the rotationally supported tori usually used in
the general-relativistic magnetohydrodynamic (GRMHD) community
\citep{Porth2019}. 

\begin{table*}
  \caption{Consequences  of the two different assumptions
        that (i) magnetic field dragging and flux-freezing
        is effective and (ii) that magnetic diffusion is dominant}
        \centering
  \begin{tabular}{l|c|c|}
    \hline
    \hline
          & flux-freezing effective& diffusion dominant\\
     \hline
	  &  MAD  &   not MAD (FLARE model) \\
  \hline
          MRI & suppressed  & active \\
          jet &  MAD jet  & no steady jet \\
          expected outflow power & huge and steady  & intermittent\\
  \hline
	  43 \& 86 GHz image & extended jet base   &  non-extended (when
	  non-flaring)\footnote{See Section \ref{sec:sgra}.} \\
          & & maybe extended when flaring  \\
\hline
\hline

  \end{tabular}
  \label{tab:initial}
\end{table*}

This novel multi-scale approach assumes ideal magnetohydrodynamics, which
has as a consequence, coherent magnetic field from large radii to be
continuously advected, to form a MAD flow in the vicinity of the black
hole \citep{Ressler2020b}.  

In what follows, we present a comprehensive description of what would be
different assuming a dominant magnetic diffusivity in the process of
accretion generated through the stellar winds. To this end, we describe
two MHD simulations, one that monitors the evolution of the magnetic
field in an accretion disk where the MRI has fully developed, and another
one where the magnetic field of different polarity is advected on the low
density region above the BH and results in flaring activity, which we
call the FLARE model. Estimations of the overall energy of the produced
flares are presented and discussed. 

This letter is organised as follows: in Section \ref{sec:dif} we discuss
in detail the different implications that follow from these assumptions,
in Section \ref{sec:mri} we present results from numerical simulations of
MRI turbulence and the magnetic field structure in the close-by accretion
disk. In Section \ref{sec:sgra} we present a GRMHD simulation (FLARE
model) that mimics the MRI turbulent state and where plasmoids are
generated above the black hole, and then compare ray-traced images from
this simulation.  Finally, we summarize and conclude in Section
\ref{sec:con}.

\section{Diffusion versus flux-freezing assumption and magnetic 
field reversals}
\label{sec:dif}

The flow generated by stellar winds around SgrA* continuously accrete
magnetic field, but as the field lines are dragged, they tend to diffuse
out. This may limit any effective process of magnetic field dragging
\citep{Heyvaerts1996}. 
As the flow and magnetic field are generated far from the BH, it needs to
be advected all the way to its vicinity. However, at distances of $300 \
- \ 10000\ r_g$ a diffusive accretion disk may form as MRI is triggered.
This argument will hold if the MRI linear growth time scale is smaller
than the viscous advection timescale, a robust estimation of this is not
easy to be calculated. 

The discussion about the efficiency of the magnetic flux-freezing
assumption is not new, also  MRI turbulent discs and the effect of
magnetic diffusion have been studied for decades. However, to interpret
EHT and GRAVITY results, such discussions are not put forward as
relevant. In this short study we highlight the importance of these
effects and propose that such long-studied effects are crucial in
interpreting SgrA$^*$ results.

In the case that turbulent viscosity is of the same order of magnetic
diffusivity, then the magnetic field is not dragged effectively. This is
set by the magnetic Reynolds number, in its simplest form $R_{\rm m} =u L
/\eta$, where $u$ is the velocity of the flow, $L$ is the length scale
and $\eta$ the resistivity/magnetic diffusivity. The value of the
magnetic Reynolds number can set the process along the way from the
ejection of the magnetized stellar winds,  all the way to the central
suppermassive BH \citep{Balbus2008}. For high $R_{\rm m}$ magnetic field
dragging is highly efficient, whereas for values closer to unity and
below this is not the case \citep{Contopoulos2015}. A transition of
$R_{\rm m}$ from low to high, if it occurs at all, will occur in the
inner disk regions $R_{\rm m}\propto \left( \frac{L}{1000R_g} \right) \left( 
\frac{u}{0.001 c}\right) << 1$ \citep{Faghei2012}. We need to note that in the
simulations reported here, no physical resistivity was used and only
numerical diffusion is present.

Thus, any flux-freezing argument will fail to produce a dynamically
important magnetic field at the vicinity of the BH.  The situation will
be different (from the MAD configuration) with a turbulent inner-region
accretion disk, which will be weakly magnetized and MRI is fully
operating from the seed poloidal field that came from the stellar winds.
Now if the magnetic field is not strong enough to suppress the MRI, the
field reversals will occur with a predetermined period. From recent
sheering box simulations we can extract that magnetic field reversals
occur in around 10 local orbital periods, where local refers to $\approx
10 R_g$ (extracted from the simulation \citep{Dhang2019a}, also see
\citep{Hogg2018}). As was pointed out previously, the accretion disk
could be MRI unstable from a radius of   $300 \ - \ 10000\ r_g$, however,
this will not set the scale of the changing of magnetic field polarity as
is shown in the simulations \citep{Dhang2019a,Hogg2018}.

This will set the length scale and the timescale of
field reversals, assuming that the local orbital period is $T_{\rm
local}= 2\pi (R_g/c)(R_{\rm local}/R_g)^{3/2}$, assuming a Keplerian
orbital period. Thus, for $R_{\rm local}=10R_g$ the corresponding period
is $T_{\rm local}\approx 200 R_g/c$, and then the corresponding timescale
for magnetic field reversals is $t_{\rm mag}\approx 2\times 10^3 R_g/c$,
which to length scale of $l_{\rm mag}\approx 10^3R_g$
\citep{Giannios2019}. This length could hint also to the transition
radius and maybe important for the particular case of SgrA*
\citep{Quataert1999,Xu2013}. It is important to note here that our
estimations are based on the simulations of \citet{Dhang2019a} and thus
the influence of the spin of the black hole is not taken into account
this could potentially affect the resulting timescale. 

In order to clearly differentiate the consequences for each
assumption we list them respectively, to show the important implications
that each assumption leads to.  At this point we take the two different
assumptions (i) dragging of magnetic field is effective and the
flux-freezing argument holds and (ii) that magnetic diffusivity is
dominant at some point of the flow through the millions of gravitational
radii that travels from the stellar winds to the BH. The discussion for
the two different cases is summarized in Table \ref{tab:initial}.

For case (i) multi-scale simulations have shown that the accretion flow
becomes magnetically arrested and a MAD state is established. The
dynamically important magnetic field suppresses the MRI which is no
longer important for the evolution of the inner-region flow
\citep{Ressler2020}.  A narrow disk-like structure is produced around the
central BH due to the impact of the strong magnetic field
\citep{Ressler2020b}. A MAD configuration is always accompanied with a
magnetized jet along the axis of the angular momentum  of the BH (for a
spinning BH, tilted disks excluded), MAD jets are expected to have huge
power and this is yet to be observed for SgrA*. This has also a straight
consequence in the compact radio image of the source \citep{Anantua2020},
which however may not be so strong in the frequency that EHT is
observing, the 230 GHz, but rather in 43 and 86 GHz where any extended
emission from the jet base with strongly affect and enlarge the image
size \citep{Davelaar2018}. 
Also a MAD configuration
has certain restrictions when applied to the whole spectrum of SgrA*, in
order to fit both the standard flux in quiescence and when flaring occurs
\citep{Chatterjee2020b}.  

The last point to touch, is about the ability of a MAD state accretion
for SgrA* to provide the observed flaring activity together with the
characteristics, like the observed orbital motion during flares as
observed by \citet{Abuter2018b}.  Reconnection events and flux eruptions
that periodically occur in MAD simulations were put in place to explain
the flares together with their detailed characteristics
\citep{Dexter2020}. The computed near-infrared (NIR) lightcurve are
promising, however the flux centroid motion from the simulation NIR
flares is much more restricted than the observed one. Moreover, the
orbiting flux tubes that can potentially explain the flares show a
sub-keplerian motion \citep{Porth2021}, whereas hot spot models of the
flux centroid motion during the flares hint to a super-keplerian motion
\citep{Matsumoto2020}. Another important observational feature that we
could potentially distinguish models is polarization, previous results
have hinted to a rather significant polarization fraction
\citep{Johnson2015}, however to analyze polarization measurements from
numerical simulations is not a trivial step \citep{Gold2017}.

\begin{figure}
    \includegraphics[width=0.46\textwidth]{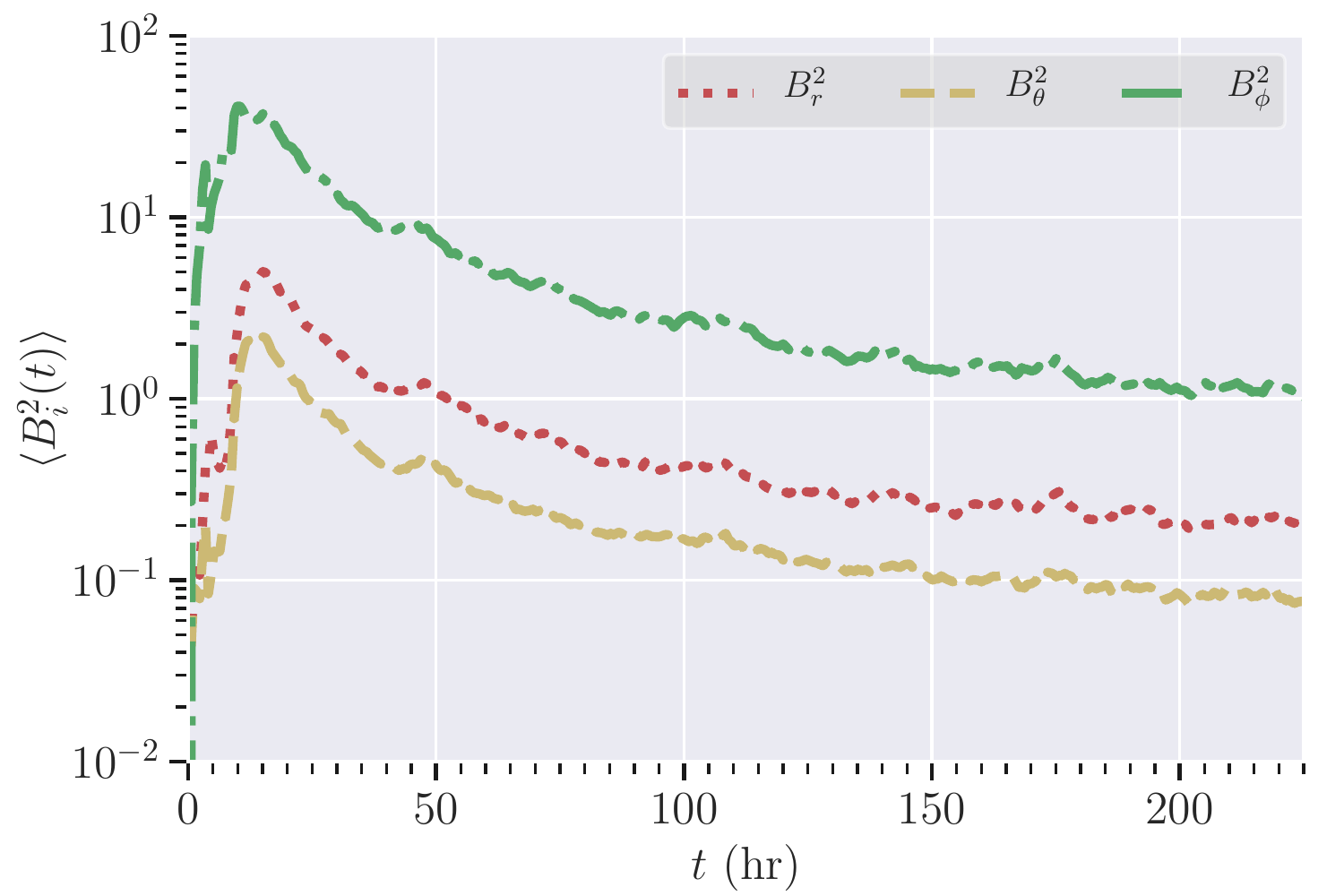}
 \caption{Time evolution of the volume averaged magnetic field energy for
        each component. Volume average is done over the region $r \in
        [6,40]$, $\theta \in [50^{\circ},130^{\circ}]$ and $\phi \in
        [0,2 \pi]$. The timescale is normalized for SgrA*.  }
        \label{fig:Bevo}
\end{figure}
\begin{figure}
    \includegraphics[width=0.46\textwidth]{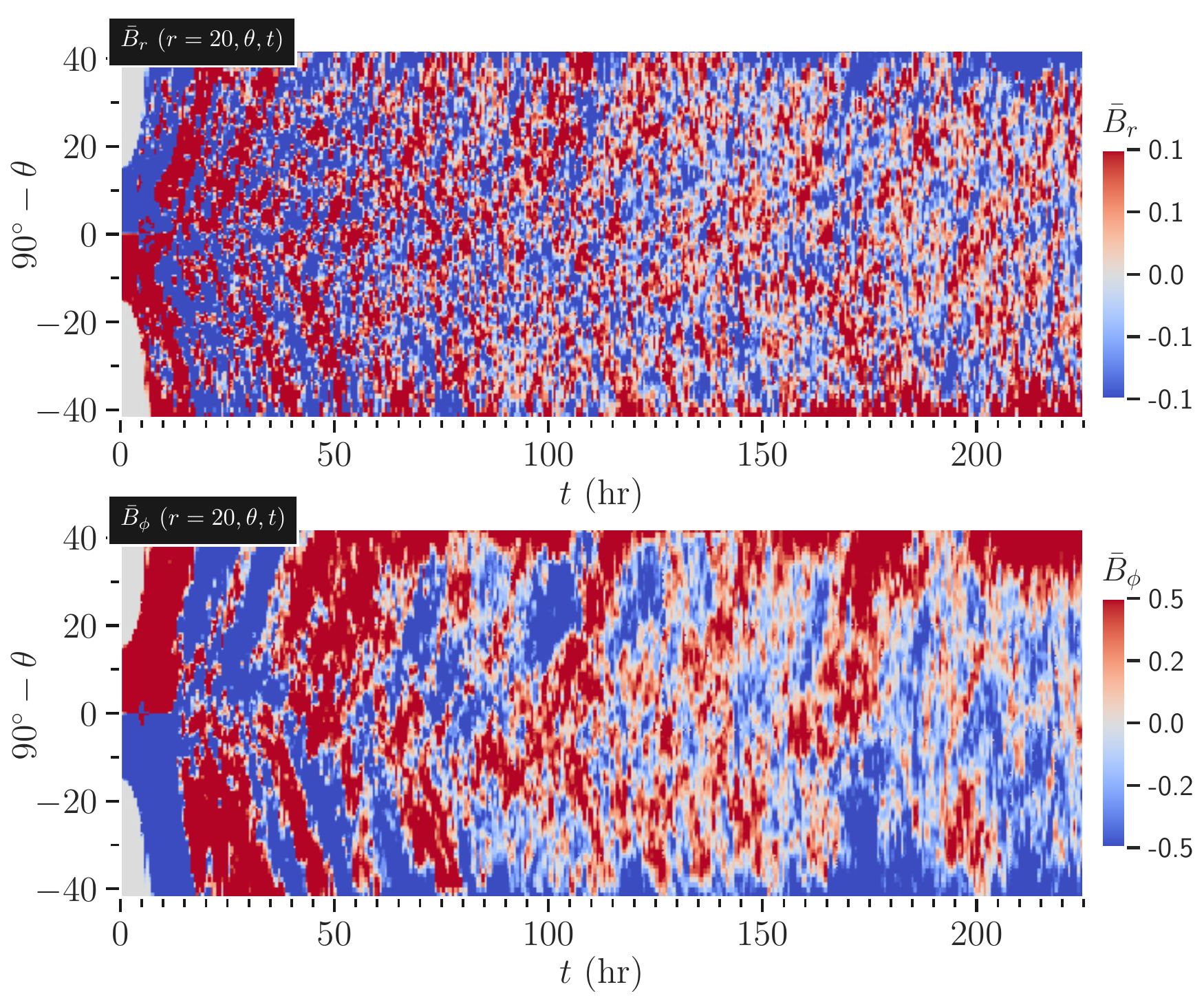}
 \caption{Butterfly diagram: Variation of azimuthally averaged radial
        ($\bar{B}_r$) and toroidal ($\bar{B}_{\phi}$) fields with
        latitude ($90^{\circ}-\theta$) and time. Time is rescaled
        appropriately for Sgr A*. } \label{fig:but}
\end{figure}

In the second case (ii), the whole picture of the inner disc accretion
flow close the central BH, is different. The magnetic field is not so
strong and MRI is not suppressed, thus is active and may dictate the
field accretion. A well-known and deeply studied phenomenon in accretion
discs with a stratification is an oscillating mean toroidal magnetic
field.  Toroidal magnetic field buoyantly rises from the mid-plane to the
upper coronal regions. This unique feature is commonly referred to as the
butterfly diagram, and is has been seen from local
\citep{Brandenburg1995,Hawley1996} but also in global simulations
\citep{Beckwith2011b,SorathiaReynolds2012}.  In the scale-free
simulations, the oscillating toroidal magnetic field changes sign
approximately every 10 local orbits, which scaled for SgrA* yields a 6-8
hour periodicity at around $\approx 20\, r_g$ from the BH. However, it
must be pointed out that unlike a geometrically thin disc, for a
geometrical thick radiatively inefficient accretion flow (RIAF), the
dynamo cycle is intermittent, both poloidal and toroidal fields flip sign
irregularly (\citealt{Hogg2018, Dhang2019a, Dhang2020}, also see Fig.
\ref{fig:but}). These dynamo generated large-scale magnetic fields tend
to be affected by the turbulent pumping (an effect which transports
large-scale magnetic field from the turbulent region to the laminar
region) resulting in transport of large-scale magnetic fields from the
turbulent disc region to the less turbulent coronal region
\citep{Dhang2020}. Hence this alternating polarity magnetic field can
reconnect either inside the high $\beta$ disk giving rise to heating of
the inner disk, or at the low density and low $\beta$ (highly magnetized)
polar region above the BH and eventually form hot spots/ plasmoids that
can become unbound and fly away from the black hole.

Reconnection in the vicinity of the black hole is operating and gives
rise to plasmoid formation
\citep{Parfrey2015,Mahlmann2020,Nathanail2020,Ripperda2020}. In the next
Section we will present results from such a simulation and give estimates
on the energy that such flares have. Such conditions could give a similar
radio image in the 230GHz, but there will be a significant difference in
the 43 \& 86 GHZ. During a quiescent state the radio image in those
frequencies will be less extended than the MAD case due to the absence of
a jet. On the other, a radio image at these observed simultaneously with
a flare, would be expected to have several larger scale features, which
would imply an extended size, maybe similar to the MAD case. In this case
models for flares accompanied with orbiting motion in horizon scales can
be expected as the ones found in GRMHD simulations \citet{Nathanail2020}
and will be also discussed in Section \ref{sec:sgra}. These models would
be similar to the plasmoid model from \citet{Ball2020}.

\section{MRI-turbulence} 
\label{sec:mri}

The problem of angular momentum transport in accretion flows and the
source of turbulence can be addressed through the action of the MRI which
was studied by \citet{Velikhov1959,Chandrasekhar1960}, but put into play
to resolve this long-standing problem by \citet{Balbus1991}. The linear
and the non-linear regimes of the MRI have been studied  in local shearing box
simulations, recently also in global simulations
\citep{Beckwith2011b,Hogg2018,Dhang2019a, Dhang2020} and GRMHD high resolution
simulations \citep{Liska2018b}. How a large-scale magnetic field is generated in
an accretion flow is still debatable. The advection of the magnetic field from
large radii has been proposed \citep{Lubow1994,Lovelace2009} or the in situ
production of a large scale magnetic field by a dynamo process
\citep{Brandenburg2011}. The production in the vicinity of the BH has been also
proposed to be dictated by a battery process that has distinctive observational
signatures \citep{Contopoulos1998},  however for an AGN with luminosity as low
as that one of SgrA* would require an extremely long timescale to build such a
field \citep{Contopoulos2015,Contopoulos2018}.

\begin{table}
  \caption{Energetics of the reported flares in Fig. \ref{fig:states1} at
        $4620\ M$ in the upper panel and at $7800\ M$ in the lower
        panel. The assumption on the accretion rate is reported in the
        first column, then the simulation time, the magnetic and total
        energy and the ratio between the magnetic field components.}
        \centering
  \begin{tabular}{l|c|c|c|c|c}
    \hline
    \hline
          $\dot{m}$  & time &  magnetic energy  & total energy &
         $Bz/B_{\phi}$   \\
          ${\rm [M_{\odot}/year]}$  &  ${\rm [M]}$  &  ${\rm [erg]}$  &${\rm   [erg]}$  &  \\
        \hline
   $10^{-7}$ & 4620 &  $8.9\times10^{38}$  & $3\times10^{39}$  & 1.14  \\
   $10^{-8}$ & 4620 &  $8.9\times10^{37}$  & $3\times10^{38}$  & 1.14  \\
   $10^{-7}$ & 7800 &  $1.4\times10^{38}$  & $5\times10^{38}$  & 1.46  \\
   $10^{-8}$ & 7800 &  $1.4\times10^{37}$  & $5\times10^{37}$  & 1.46  \\

          \hline
  \end{tabular}
  \label{tab:2}
\end{table}

We are particularly interested in the generation of a large scale
magnetic field by the MRI driven dynamo in a  weakly magnetised RIAF
\citep{Beckwith2011b}.  Here we will describe the results from the global
3D MHD simulation that investigates such generation of magnetic fields
and the self-sustained dynamo process in a geometrically thick
radiatively inefficient (RIAF) accretion disk (model M-2P of
\cite{Dhang2019a}). The simulation has an inner/ outer radius of $4R_g/
140R_g$ respectively and a resolution of ($r, \theta, \phi)=(296,128,
512)$ It is important to note that to have such a high
resolution in the disk and also close to the event horizon is costly and
forbiden for present simulations, for this reason we employ one
simulation that focuses in the disk evolution and another that focuses on
the black hole activity.  The simulation is initialised to an equilibrium
torus \citep{Papaloizou84}  threaded by the poloidal magnetic field loops
of plasma $\beta \sim 800$, such that the magnetically, RIAF remains in
the regime of standard and normal evolution (SANE) type of accretion disk
in the quasi-stationary state.

Fig. \ref{fig:Bevo} shows the evolution of the magnetic energy in the
accretion flow.  Time evolution follows this path: initially both poloidal
($B_r$ and $B_{\theta}$)  and toroidal ($B_{\phi}$) components of the magnetic
field grow exponentially.  The poloidal components are amplified by the linear
MRI, while shear generates the toroidal components with different sign above
and below the equatorial plane. In due time the system enters the non-linear
regime and parasitic instabilities become important  \citep{Goodman1994},
leading to the full development of turbulence throughout the accretion disk.
Here it is to be noted that the initial zero toroidal magnetic component
evolves in time to increase significantly comprising of almost eighty percent
of the total magnetic energy in the quasi-steady state.

Fig. \ref{fig:but} shows the evolution of the azimuthally averaged radial
($\bar{B}_r$) and toroidal ($\bar{B}_{\phi}$) magnetic fields with latitude
($90^{\circ}-\theta$) and time at a radius $r=20 \ r_g$.  While the strong
shear transforms the poloidal fields into the toroidal fields, the poloidal
fields can be regenerated by an $\alpha$-effect  out of the toroidal fields
\citep{Brandenburg1995,Dhang2019a,Dhang2020}.  This cycle goes on resulting in
large-scale magnetic fields (both poloidal and toroidal) of alternate polarity
in time and develops a butterfly diagram.  This is a phenomenon widely known
and found in local and global simulations of accretion disks that study the
MRI, moreover several studies hint to a certain periodicity, which for SgrA*
would be a matter of hours
\citep{Brandenburg1995,Hawley1996,SorathiaReynolds2012,Hogg2018,Dhang2019a}.

\begin{figure} \begin{center}
        \includegraphics[width=0.45\textwidth]{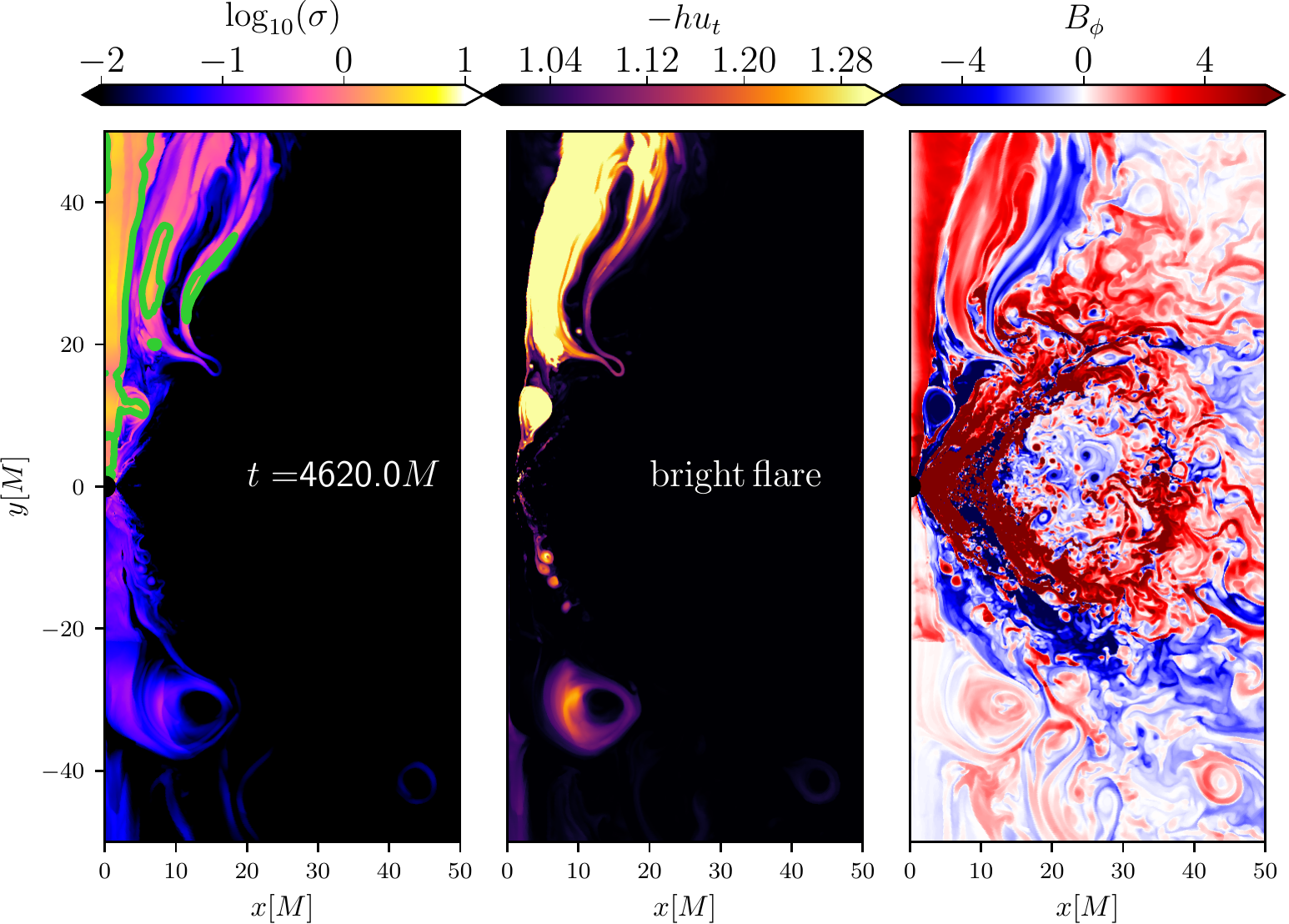}
    \includegraphics[width=0.45\textwidth]{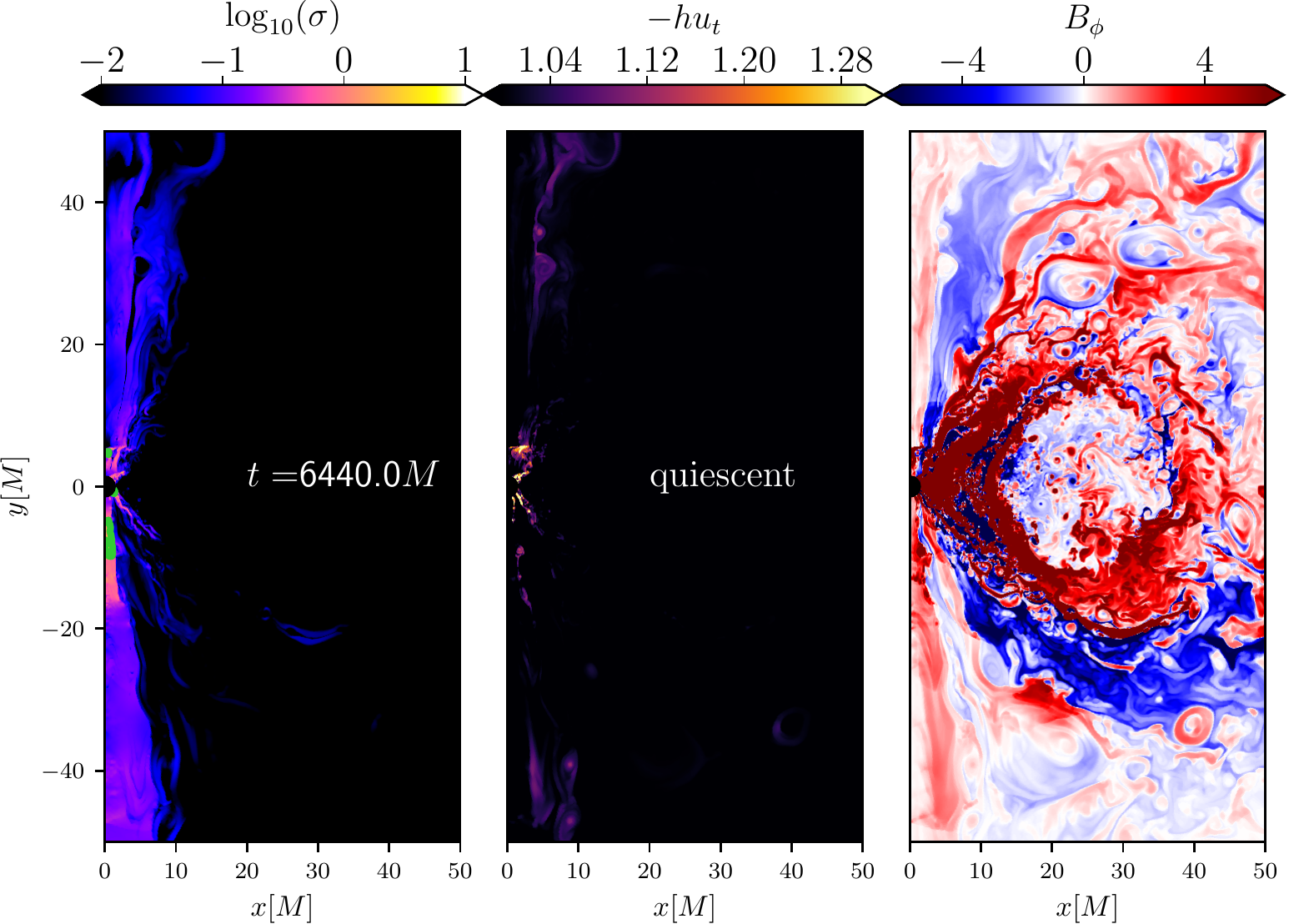}
    \includegraphics[width=0.45\textwidth]{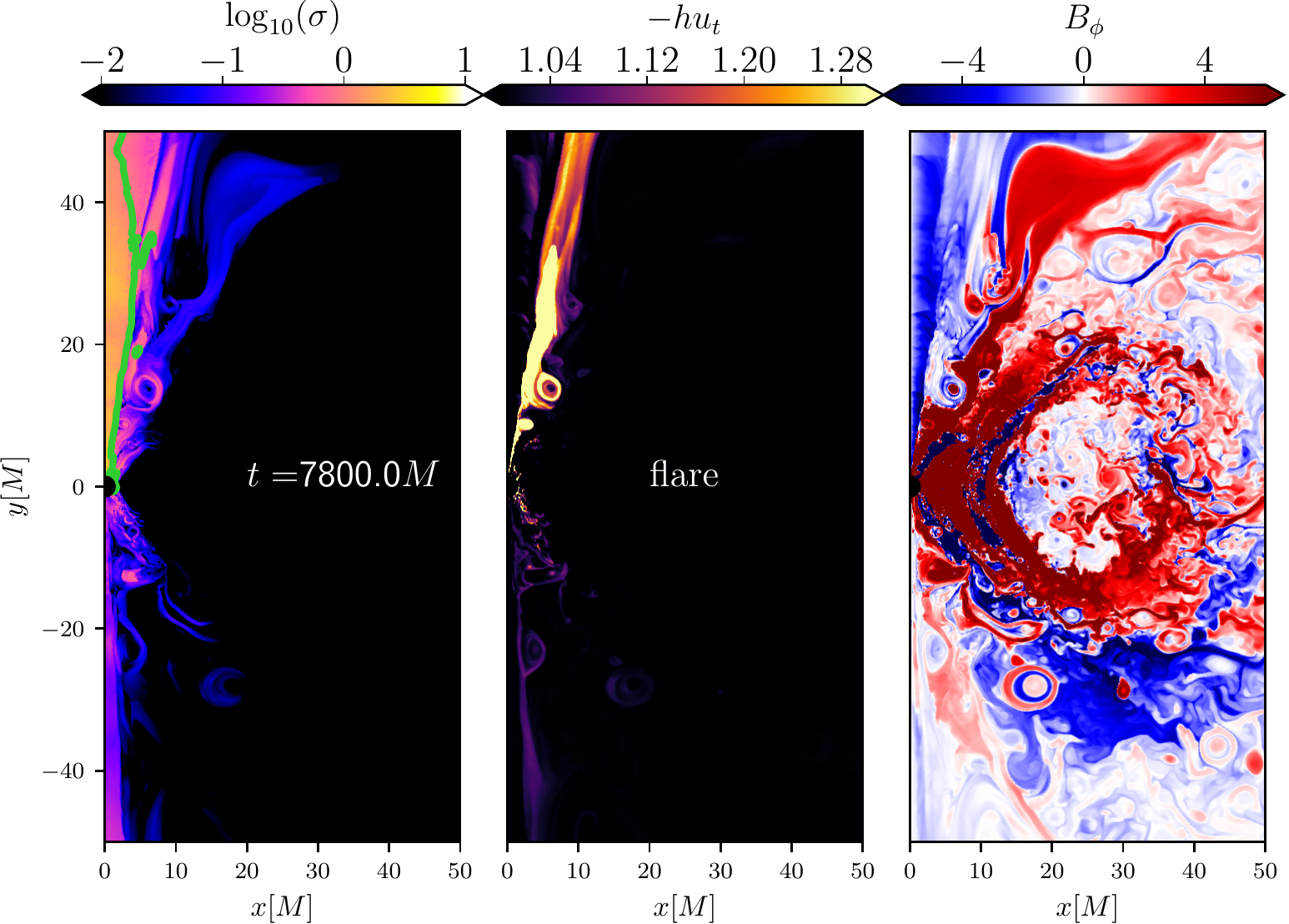}
\end{center} \caption{Different states during the evolution of the
accretion flow with a bright flare (upper  panel), a state of
quiescence (middle panel) and another flare (lower panel). In
both flares  an orbiting plasmoid/hot spot can be identified.
The time lapse from the upper to the middle panel is 10 hours and
from the middle to the lower 7.5 hours (for SgrA* scales). The
first column shows the  magnetization $\sigma$, the middle column
the criterion for unbound matter $-hu_t$ (where $-hu_t>1$ means
unbound) and the last column the toroidal magnetic field $B_{\phi}$.
The reported snapshots are from the 2D FLARE simulation.}
\label{fig:states1}
\end{figure}

\begin{figure*}
    \includegraphics[width=0.93\textwidth]{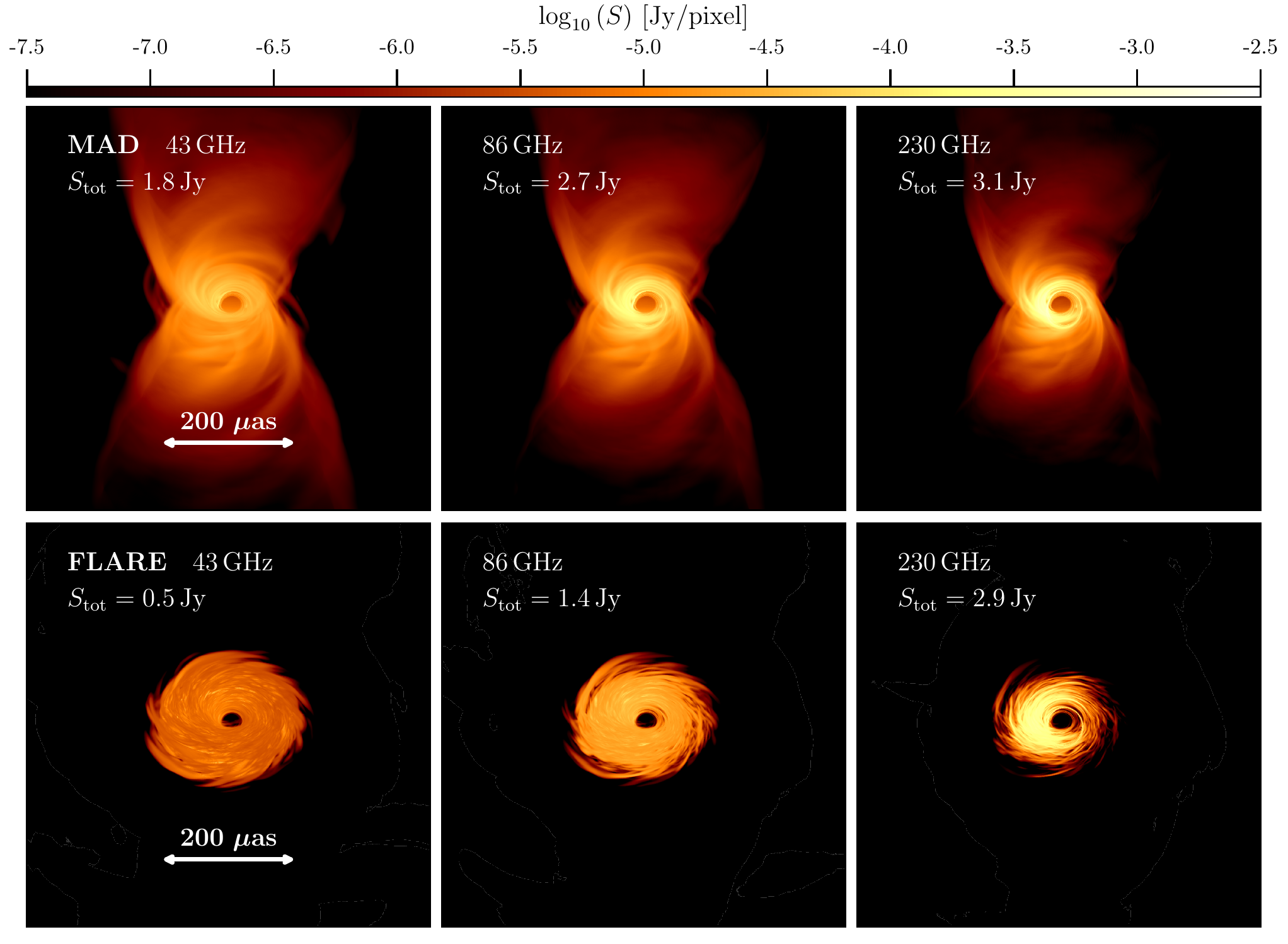}
        \caption{GRRT images for SgrA$^\star$ computed for three 
	different frequencies $43$\,GHz (left column), 
	$86$\,GHz (middle column) and $230 {\rm GHz}$ (right
	column). The top row displays the images generated from a MAD
	GRMHD model with spin, a$^\star=0.9375$
	\citep{Osorio2022,Fromm2021} and the bottom row shows images
	obtained from a 3D FLARE model with 
	spin $a^\star=0.9375$ \citep{Nathanail2021b}.}
        \label{fig:im}
\end{figure*}

A strong turbulent pumping is present in the MRI driven dynamo  giving rise
to transport of the large-scale fields from the turbulent region close to the
mid-plane to the less turbulent coronal region \citep{Dhang2020}.
Accretion of the large-scale magnetic fields is found to be more
efficient in the coronal region compared to that in the disc \citep{Guilet2013}.
As these alternating polarity magnetic fields are advected onto the black
hole, they reconnect and produce structures, such as
plasmoids. In some cases the plasmoids fall onto the BH but they
may also become energised and eventually become unbound and responsible
for flaring activity.

\section{The Different states of $\rm SgrA^*$ and image differences} 
\label{sec:sgra}

In this section we first discuss how the accretion of the butterfly
diagram onto the BH may give rise to distinctive flaring features and
present a qualitative description of the results discussed thoroughly in
\citep{Nathanail2020,Nathanail2021b} called the FLARE model, and then show
results of ray traced images in general relativity (GRRT) in order to
show the size of the image changes for MAD models and the FLARE models we
have discussed here.  These features can account for the different states
observed in the compact region at the center of our galaxy, $\rm SgrA^*$.
We report high resolution, 2D \citep{Nathanail2020} and 3D
\citep{Nathanail2021b} GRMHD simulations performed with the \bhac code
\cite{Porth2017}, a logarithmic grid is employed with $4096 \times  2048$
in $r$ and $\theta$ for the 2D simulation, whereas an effective
resolution of  $1050 \times 768 \times 384$ in $r,\ \theta$ and $\phi$
for the 3D, an outer boundary at $2500\, r_g$ and $500\, r_g$
respectively, and a Kerr spacetime with a dimensionless spin parameter of
$a=0.9375$ for both 2D and 3D. The inner and outer radii of the torus are
at $r_{\rm in} = 6\, r_g$ and  $r_{\rm out} = 12\, r_g$.  When the simulation reaches a
steady state the average normalized magnetic flux at the horizon is
$\Phi_{BH}/ \sqrt{\dot{m}} \approx 0.5$.  This magnetic flux fluctuates,
as the loops anninhilate, but never approach the MAD value.  The setup
for the MAD simulation has an effective resolution of $384^3$ in $r,\
\theta$ and $\phi$, an outer boundary at $2500\, r_g$, the same Kerr
spacetime and inner and outer radii of the torus are at $r_{\rm in} = 20\, r_g$
and $r_{\rm out} = 41\, r_g$. Apart from the disk size, the major difference
between MAD and FLARE simulations are the initial magnetic field
configuration, for MAD this is a nested loop embedded in the initial
torus and for FLARE the initial torus configuration is seeded with loops
of alternating polarity.

In the FLARE model loops of alternating polarity are advected with the
flow to the BH. This setup does not yield a persistent jet structure, but
may produce striped jets \citep{Giannios2019}.  The evolution of such an
accretion flow together with its characteristics was presented in detail
in \cite{Nathanail2020,Nathanail2021b}.  Here, we restrict into a
consideration of these simulation  with a comparison to the activity of
$\rm SgrA^*$. Our focus is to find, during the evolution of the accretion
disk and the advection of the magnetic loops, times that the whole system
is radically changing behavior. At the point that two layers of different
magnetic field polarity merge in the low density region above the BH,
plasma gets energy and forms plasmoids through magnetic reconnection,
these features become unbound and leave the BH in a spiral outward orbit,
as seen in the upper panel of Fig. \ref{fig:states1}.  

After a major reconnection event, we usually observe a quiescent state
for some time (middle panel of Fig. \ref{fig:states1}), till the flow
builds a considerable magnetic flux of one polarity. This structure is
subsequently annihilated by the advection of magnetic flux of different
polarity and another flare (unbound plasmoids, identified through a
cutoff in magnetization) is generated, as in the lower panel of Fig.
\ref{fig:states1}, with a recurring timescale of $\approx 1200 \ {\rm
M}$. Flare from flare may differ in overall energy, which depends on the
amount of magnetic flux that reconnects at the reconnection site. To
estimate the energy of each flare we assume a mass accretion rate of
$10^{-8}\ -\  10^{-7}\ M_{\odot}/year$, for the two boundary values of
this range we estimate the energy in the flares and report them in Table
\ref{tab:2}. Since our simulations are 2D, we assume an angular size of
$\pi/10$ for the plasmoids. The resulting energetics acquire enough
energy to explain bright flares from SgrA* \citep{Bouffard2019}. Another
important point is that the magnetic field in the flares is predominantly
poloidal, as seen in the last column of Table \ref{tab:2}, and thus could
explain the linear polarisation reported \citep{Abuter2018b}, note that
this does not hold inside the dense accretion disk where the field is
mostly toroidal due to the MRI and the differential rotation.

To further compare the standard MAD GRMHD simulations \citep[see,
e.g.,][]{Osorio2022,Fromm2021} with the ones reported in this work we
performed general relativistic radiative transfer (GRRT) calculations.
Given that the GRMHD simulations are scale free (except for the spin and
initial magnetic field topology) we adjusted to the galactic centre
during the GRRT by setting the mass and distance. We used a black hole
mass of $4.14\times10^6\,M_{\sun}$ and a distance of 8.14\,kpc
\citep{Gravity2019}. We iterate the mass accretion rate\footnote{The
resulting mass accretion rate is consistent with the values used in Table
\ref{tab:2}.} until we obtain a flux of $\sim$3\,Jy at 230\,GHz in
agreement with the SgrA$^\star$ observations of \citet{Bower2019}.
During the radiative transfer we exclude regions in the jet funnel with
large magnetisation $\sigma>1$ to avoid contamination from floor values
introduced during the simulations to ensure numerical stability. The
radiative calculation is performed using the well tested GRRT code
\textit{BHOSS} \citep{Younsi2012,Younsi2020}. Since your GRMHD
simulations do not evolve the radiating electrons we need to reconstruct
their properties, e.g., temperature and number density, from the ions.
Therefore, we applied the so-called $R-\beta$ model
\citep{Moscibrodzka2016} relating the ion temperature, $T_{\rm i}$
together with the plasma beta, $\beta=p_{\rm gas}/p_{\rm mag}$ via a free
parameter labeled as $R_{\rm high}$ to the electron temperature:

\begin{equation}
\Theta_{\rm e}= \frac{p m_{\rm p}/m_{\rm e} }{ \rho T_{\rm ratio}},\,  
	\quad T_{\rm ratio} \equiv \frac{T_{\rm p}}{T_{\rm e}}=
	\frac{1+R_{\rm high}\beta^{2}}{1+\beta^{2}},
\label{eq:Te}
\end{equation}
with $m_{\rm p}$ and $m_{\rm e}$  the proton and electron masses,
respectively. During the GRRT we assumed a thermal distribution electrons
while we fixed the inclination angle to 50$^\circ$ and used a $R_{\rm
high}=5$\footnote{The values are chosen to highlight the launched jet in
the MAD models}. The result of the radiative transport can be seen in
Fig.~\ref{fig:im}. The differences between the MAD model (top row) and
the FLARE model (bottom row) are clearly visible and most striking one is
the missing jet feature in the FLARE images. Another feature of the MAD
model are bright arcs, connected to flux tubes anchored mainly in the
accretion disk scale \citep{Porth2021}. In contrast, the FLARE images do
not show a permanent jet\footnote{as mentioned above jets are produced as
transient feature in the FLARE model}. The flux is concentrated to the
innermost disk regions  $r\sim100\,\rm{\mu as}$ with a steep decay with
decreasing frequency as compared to the MAD model. Given these
significant differences in the images and in the spectral behaviour the
two models should be distinguishable in recent and future observations of
the Event Horizon Telescope.

\section{Conclusions} 
\label{sec:con}

In this letter we stress out the importance of using well-studied MRI
turbulent discs for the interpretation of EHT and GRAVITY results. We
discuss the possibility that the magnetic field structure in the vicinity
of the super massive BH at the center of our galaxy, SgrA*, may be
totally dictated by the action of the MRI. 
This would have as a consequence that the inner accretion flow will never
become magnetically arrested (MAD).  We present in detail the
different ways to distinguish the two scenarios, mainly by the
images in 43 and 86 GHz that will uncover an extended structure at  the
base, close to the BH, but also the polarization signatures. These can be
tested by near future observations.

Results from global simulations of the MRI in accretion disks, provide
evidence for the action of the MRI and its impact to the magnetic field
geometry showing an alternating polarity in an 6-8 hours periodicity.
When poloidal field of different polarity is accreted, it reconnects and
provides an energy release to produce plasmoid/hot spots that can further
be observed as flares for SgrA*.

Lastly, we describe a specific GRMHD simulation that tries to mimic this
behavior for the production of plasmoids interchanging with states of
quiescence in the vicinity of the BH and we estimate the magnetic energy
in the  range of $\approx 10^{35 - 38}\ {\rm erg}$ for the flares,
assuming a decent mass accretion rate of $\approx 10^{-8}\
M_{\odot}/year$ for SgrA*, and show that the intrinsic poloidal magentic
field in these flares can also explain the observed polarization imprint
\citep{Abuter2018b}.

\bigskip
\noindent\textit{\textbf{Acknowledgements.~}}
The simulations were performed  on GOETHE at CSC-Frankfurt.  AN was
supported by the Hellenic Foundation for Research and Innovation
(H.F.R.I.) under the 2nd Call for H.F.R.I. Research Projects to support
Post-Doctoral Researchers (Project Number: 00634) This research is supported by the DFG research grant ``Jet physics on horizon scales and beyond" (Grant No.  FR 4069/2-1) and the ERC synergy grant ``BlackHoleCam: Imaging the Event Horizon of Black Holes" (Grant No. 610058)  \\
\noindent\textit{\textbf{Data Availability.~}}
The data underlying this article will be shared on reasonable request to
the corresponding author.




\section*{}


\bibliographystyle{mnras}
\bibliography{aeireferences}



%
%
%

\end{document}